# A macroscopic model that connects the molar excess entropy of a supercooled liquid near its glass transition temperature to its viscosity

## Hiroshi Matsuoka[a)]

*Department of Physics, Illinois State University, Normal, Illinois, 61790-4560, USA*

For a deeply supercooled liquid near its glass transition temperature, we suggest a possible way to connect the temperature dependence of its molar excess entropy to that of its viscosity by constructing a macroscopic model, where the deeply supercooled liquid is assumed to be a mixture of solid-like and liquid-like micro regions. In this model, we assume that the mole fraction $x$ of the liquid-like micro regions tends to zero as the temperature $T$ of the liquid is decreased and extrapolated to a temperature $T_g^*$, which we assume to be below but close to the lowest glass transition temperature $T_g$ attainable with the slowest possible cooling rate for the liquid. Without referring to any specific microscopic nature of the solid-like and liquid-like micro regions, we also assume that near $T_g$, the molar enthalpy of the solid-like micro regions is lower than that of the liquid-like micro regions. We then show that the temperature dependence of $x$ is directly related to that of the molar excess entropy. Close to $T_g$, we assume that an activated motion of the solid-like micro regions controls the viscosity and that this activated motion is a collective motion involving practically all of the solid-like micro-regions so that the molar activation free energy $\Delta g_a$ for the activated motion is proportional to the mole fraction, $1 - x$, of the solid-like micro regions. The temperature dependence of the viscosity is thus connected to that of the molar excess entropy $s_e$ through the temperature dependence of the mole fraction $x$.

---

[a)] Electronic mail: hmb@phy.ilstu.edu

As an example, we apply our model to a class of glass formers for which $s_e$ at temperatures near $T_g$ is well approximated by $s_e \propto 1 - T_K/T$ with $T_K < T_g \equiv T_g^*$ and find their viscosities to be well approximated by the Vogel-Fulcher-Tamman equation for temperatures very close to $T_g$. We also find that a parameter $a$ appearing in the temperature dependence of $x$ for a glass former in this class is a measure for its fragility. As this class includes both fragile and strong glass formers, our model applies to both fragile and strong glass formers. We estimate the values of three parameters in our model for three glass formers in this class, $o$-terphenyl, 3-bromopentane, and $Pd_{40}Ni_{40}P_{20}$, which is the least fragile among these three. Finally, we also suggest a way to test our assumption about the solid-like and liquid-like micro regions by means of molecular dynamics simulations of model liquids.

## I. INTRODUCTION AND A SUMMARY OF OUR RESULTS

Being cooled at sufficiently fast rates, many liquids can be supercooled below their melting temperatures without becoming crystalline solids until their viscosities become so high (*e.g.*, $10^{13}$ P) that they appear to behave like solids, which are called glasses. The changeover from a non-equilibrium supercooled liquid state to a non-equilibrium glass state proceeds continuously around a glass transition temperature, which can be defined as the temperature at which the viscosity of the supercooled liquid reaches some chosen value such as $10^{13}$ P. The glass transition occurs at a lower temperature for a slower cooling rate. In this article, $T_g$ denotes the lowest glass transition temperature attainable with the slowest possible cooling rate.

Despite many decades of experimental and theoretical work, there is no consensus on the nature of the glass transition[1]. However, many deeply supercooled liquids near their glass transition temperatures exhibit the following two properties: (i) the molar excess entropy $s_e$ of each supercooled liquid decreases smoothly as its temperature $T$ is decreased until it reaches $T_g$, below which $s_e$ deviates from the smooth trend it follows above $T_g$ and continues to decrease with a much less steep slope; (ii) the relaxation time $\tau$ or the viscosity $\eta$ of each supercooled liquids rapidly increases as $T$ is decreased toward $T_g$.

A natural question is whether the decrease in $s_e$ and the increase in $\tau$ or $\eta$ are related with each other. The main goal of this article is to suggest a possible way to connect these two properties by constructing a macroscopic model for a deeply supercooled liquid near $T_g$, which we assume to be a mixture of solid-like and liquid-like micro regions. We will show that these two properties are connected through the temperature dependence of the mole fraction $x$ of the liquid-like micro regions.

In our model, we assume that the mole fraction $x$ of the liquid-like micro regions tends to zero as the temperature $T$ of the liquid is decreased and extrapolated to a temperature $T_g^*$, which we assume to be below but close to the lowest glass transition temperature $T_g$. Without referring to any specific microscopic nature of the solid-like and liquid-like micro regions, we also assume that near $T_g$, the molar enthalpy $h_g$ of the solid-like micro regions is lower than the molar enthalpy $h_l$ of the liquid-like micro regions and that the molar enthalpy difference defined by $\Delta h \equiv h_l - h_g$ is almost independent of temperature over a narrow temperature range above $T_g$.

In Sec. III.G, we will find the temperature dependence of the mole fraction $x$ of the liquid-like micro regions to be directly related to that of the molar excess entropy $s_e$ by Eq. (3.24):

$$\left( \frac{\partial x}{\partial T} \right)_P = \frac{T}{\Delta h} \left( \frac{\partial s_e}{\partial T} \right)_P \qquad (1.1)$$

In Sec. III.H, we assume that close to $T_g$, an activated motion of the solid-like micro regions controls the viscosity $\eta$ of the supercooled liquid. We also assume that this activated motion is a collective motion involving practically all of the solid-like micro-regions so that the molar activation free energy $\Delta g_a$ for the activated motion is proportional to the mole fraction, $1 - x$, of the solid-like micro regions. We will then obtain Eq. (3.25):

$$\eta = \eta_0 \exp\left( \frac{\Delta g_a}{RT} \right) = \eta_0 \exp\left[ \frac{\Delta g_0 (1 - x)}{RT} \right], \qquad (1.2)$$

where $R$ is the universal gas constant and $\Delta g_0$ is a constant specific to a particular liquid. The temperature dependence of the viscosity $\eta$ is thus connected to that of the molar excess entropy $s_e$ through the temperature dependence of the mole fraction $x$.

Using Eq. (1.1) and Eq. (1.2), we will also find the molar excess heat capacity at constant pressure $c_P^e$ of the liquid to be related to the viscosity $\eta$ by

$$c_P^e = T\left(\frac{\partial s_c}{\partial T}\right)_P = \Delta h\left(\frac{\partial x}{\partial T}\right)_P = -\frac{R\Delta h}{\Delta g_0}\left[\frac{\partial}{\partial T}\left\{T\ln\left(\frac{\eta}{\eta_0}\right)\right\}\right]_P. \qquad (1.3)$$

We can therefore test our model for a particular glass former by finding whether its experimental data for $c_P^e$ and $\eta$ are consistent with this equation with two adjustable parameters, $\eta_0$ and $R\Delta h/\Delta g_0$.

Our model is somewhat similar to the Adam-Gibbs theory[2], where the temperature dependence of the relaxation time or viscosity $\eta$ is directly controlled by the temperature dependence of the molar configurational entropy $s_c$ of the supercooled liquid:

$$\eta = \eta_0 \exp\left(\frac{A}{Ts_c}\right), \qquad (1.4)$$

where $\eta_0$ and $A$ are constants specific to a particular liquid. The molar configurational entropy $s_c$ is a part of the molar excess entropy $s_e$ and decreases as the temperature of the supercooled liquid is decreased toward $T_g$ so that this equation provides a link between the decrease in $s_e$ and the increase in $\eta$.

In the Adam-Gibbs theory, a supercooled liquid is assumed to be divided into cooperatively rearranging regions with a linear size $\xi$. The molar activation free energy $\Delta_a$ for the activated motion of each region is then assumed to be proportional to the volume $\xi^3$ of the region so that $\Delta_a \propto \xi^3$ while $s_c$ is proportional to the number of these regions, which is inversely proportional to $\xi^3$ so that $s_c \propto 1/\xi^3$. Eq. (1.4) then follows from

$$\eta = \eta_0 \exp\left(\frac{\Delta_a}{RT}\right). \qquad (1.5)$$

Our model and the Adam-Gibbs theory are different mainly in the following two respects. In our model, as the temperature of the supercooled liquid approaches $T_g$, the molar activation free energy $\Delta g_a$ increases because of the increase in the mole fraction, $1 - x$, of the solid-like micro regions, while in the Adam-Gibbs theory, the molar activation free energy $\Delta_a$ increases because of the increase in the volume $\xi^3$ of each cooperatively rearranging region. In our model, the temperature dependence of the viscosity is connected to that of the molar excess entropy $s_e$, which we can obtain directly from experimental data for the molar excess heat capacity at constant pressure $c_P^e$, while in the Adam-Gibbs theory, the viscosity is controlled by the molar configurational entropy $s_c$, which is not directly accessible by experiments as it is a part of $s_e$.

In Sec. IV, we will apply our model to a class of glass formers for which $s_e$ at temperatures near $T_g$ is well approximated by the following equation[3]:

$$\frac{s_e(T)}{s_\infty} = 1 - \frac{T_K}{T} \qquad \text{(for } T \geq T_g > T_K\text{)}, \qquad (1.6)$$

where both $s_\infty$ and $T_K$, the Kauzmann temperature, are constants specific to a particular liquid. We will also assume that $T_g^*$ is very close to $T_g$ so that we can approximate $T_g^*$ by $T_g$.

Using Eq. (1.6) in Eq. (1.1), we will find the following temperature dependence of the mole fraction $x$ of the liquid-like micro regions:

$$x = a \ln\left(\frac{T}{T_g}\right), \qquad (1.7)$$

where $a$ is defined by

$$a \equiv \frac{s_\infty T_K}{\Delta h}. \qquad (1.8)$$

Using Eq. (1.7) in Eq. (1.2), we will find the viscosity very close to $T_g$ is well approximated by the following Vogel-Fulcher-Tamman equation:

$$\eta = \eta_0 \exp\left(\frac{D T_{VFT}}{T - T_{VFT}}\right) = \eta_0 \exp\left(\frac{D}{T/T_{VFT} - 1}\right), \qquad (1.9)$$

where $T_{VFT}$ and $D$ are given by

$$T_{VFT} = \frac{a}{a+1} T_g \qquad (1.10)$$

and

$$D = \frac{\Delta g_0}{R T_g a}. \qquad (1.11)$$

Note that this approximation by the Vogel-Fulcher-Tamman equation works only for temperatures very close to $T_g$. The experimental data for $\tau$ or $\eta$ for the glass formers in this class can be fit by a number of equations other than the Vogel-Fulcher-Tamman equation[4, 5]. However, for temperatures very close to $T_g$, a fit by any of these equations is practically indistinguishable from that by the Vogel-Fulcher-Tamman equation so that very close to $T_g$, the temperature dependence of $\eta$ predicted by our model is also well approximated by any of these equations.

For many liquids, especially for fragile glass formers, $T_{VFT}$ has been found to be close to $T_K$, but for some liquids, especially for some strong glass formers, $T_{VFT}$ has been found to be considerably lower than $T_K$[6]. In our model, $T_{VFT}$ and $T_K$ do not need to be equal or close to each other so that our model applies to both fragile and strong glass formers. As we will show in Sec. IV.E, if $T_g s_e(T_g)/\Delta h \approx 1$, then $T_K \approx T_{VFT}$ whereas if $T_g s_e(T_g)/\Delta h < 1$, then $T_K > T_{VFT}$.

In the Adam-Gibbs theory, if the molar configurational entropy $s_c$ of a glass former is proportional to its molar excess entropy $s_e$ so that $s_e = s_\infty^c (1 - T_K/T)$, then using this equation in Eq. (1.4), we also find the viscosity $\eta$ to follow the Vogel-Fulcher-Tamman equation:

$$\eta = \eta_0 \exp\left[\frac{A}{s_\infty^c (T - T_K)}\right].  \qquad (1.12)$$

Comparing this equation with Eq. (1.9), we find $T_{VFT} = T_K$ and $D = A/(T_K s_\infty^c)$. Generally, $s_c$ of any glass former is smaller than $s_e$ so that for a strong glass former with $T_{VFT} < T_K$, $s_c$ may decrease, as the temperature approaches $T_g$, less steeply than $s_e$ so that it may satisfy $s_e = s_\infty^c (1 - T_{VFT}/T)$. The Adam-Gibbs theory then lead to the Vogel-Fulcher-Tamman equation, Eq. (1.9), for the strong glass former.

According to Eq. (1.10), as the parameter $a$ approaches zero, $T_{VFT}$ also approaches zero so that the Vogel-Fulcher-Tamman equation approaches the Arrhenius equation, which implies that the parameter $a$ is a measure for the fragility of a supercooled liquid: the smaller the parameter $a$ is, the less fragile the supercooled liquid is.

Eq. (1.11) also shows that the strength parameter $D$, one of frequently used measures for the fragility of a supercooled liquid, depends on the parameter $a$ and $\Delta g_0/(RT_g)$ and in Sec. IV.D, we will show that the steepness index $m$, another frequently used measure for the fragility, depends on $a$ and $\Delta g_0/(RT_g)$. If $\Delta g_0/(RT_g)$ is almost constant, as is the case for three glass formers in Table I, then we find $m$ to be roughly proportional to $a + 1$ and $D$ to be roughly inversely proportional to $a$.

According to Eq. (1.7), the larger the parameter $a$ is, the more rapidly the mole fraction $x$ of the liquid-like micro regions increases as the temperature of a supercooled liquid is increased. This implies that the solid-like micro regions inside a fragile glass former are more easily destroyed by an increase in the temperature than those in a less fragile glass former. In other words, the solid-like micro regions inside a fragile glass former are "fragile" against an increase in the temperature. The parameter $a$ is therefore a measure for the fragility of the solid-like micro regions inside a supercooled liquid very close to $T_g$.

Our macroscopic model applied to this class of glass formers contains three parameters: the parameter $a$, $\Delta h$, and $\Delta g_0 / (RT_g)$. For a particular glass former, we can estimate their values from the values of $D$, $T_g / T_{VFT}$, and $s_\infty T_K$.

In Table I, we will list the estimated values of the parameter $a$, $\Delta h$, and $\Delta g_0 / (RT_g)$ for three glass formers in this class: $o$-terphenyl, 3-bromopentane, and $Pd_{40}Ni_{40}P_{20}$. Among these three, $o$-terphenyl is the most fragile while $Pd_{40}Ni_{40}P_{20}$ is the least fragile.

Finally, in Sec. VI, we will suggest a way to test our assumption about the solid-like and liquid-like micro regions by means of molecular dynamics simulations of model liquids.

## II. A DEEPLY SUPERCOOLED LIQUID AS A MIXTURE OF SOLID-LIKE AND LIQUID-LIKE MICRO REGIONS

### A. Our physical picture

Although our macroscopic model does not depend on any specific detail of a physical picture of a deeply supercooled liquid on the microscopic level, it is motivated by the following physical picture of the deeply supercooled liquid.

As the temperature of the supercooled liquid approaches $T_g$, more and more solid-like micro regions begin to appear throughout the entire liquid region. These solid-like micro regions remain microscopic in size since they cannot fill the space without creating empty gaps between them as is the case, for example, with multi-layer icosahedra each with a few layers in a single component system[12].

Near $T_g$, as the solid-like micro regions start to form a randomly connected network, the liquid region becomes disjointed and filamentary so that it is divided into liquid-like micro regions to fill the space between the solid-like micro regions.

Close to $T_g$, we assume that these solid-like micro regions have, on average, a slightly higher density than the liquid-like micro region so that, on the macroscopic level, the molar volume $v_g$ of the solid-like micro regions are lower than that of the liquid-like micro regions.

It is not entirely clear whether the molar internal energy $u_g$ of solid-like micro regions is discernibly lower than the molar internal energy $u_l$ of the liquid-like micro region although we do not expect $u_g$ to be higher than $u_l$.

However, to construct our macroscopic model, we will only need to assume that the molar enthalpy of the solid-like micro regions is lower than that of the liquid-like micro regions, which is still valid as long as there is a molar volume difference between these two types of micro regions, even if no discernible difference in the molar internal energy exists between the two.

**B. The mole fraction $x$ of the liquid-like micro regions**

In our model, the pressure $P$ and the mole number $n$ of a supercooled liquid are kept constant. The mole fraction $x$ of the liquid-like micro regions is then a function of the temperature $T$ of the supercooled liquid: $x = x(T)$. We also assume that $x$ tends to zero as $T$ is decreased and extrapolated to a temperature $T_g^*$, which we assume to be below but close to the lowest glass transition temperature $T_g$ attainable with the slowest possible cooling rate for the liquid:

$$\lim_{T \to T_g^*} x(T) = 0 \qquad\qquad (2.1)$$

and

$$\left(\frac{\partial x}{\partial T}\right)_P > 0 \,. \tag{2.2}$$

## III. MACROSCOPIC MODEL THAT CONNECTS THE MOLAR EXCESS ENTROPY TO THE VISCOSITY

To describe non-equilibrium states of a supercooled liquid, we cannot use quantities defined only for equilibrium states. In this section, we will define the macroscopic quantities that are well-defined for these non-equilibrium states and find them to be all related to the mole fraction $x$ of the liquid-like micro regions. In our model, the pressure $P$ and the mole number $n$ of the supercooled liquid are kept constant so that these macroscopic quantities are functions of the temperature $T$ of the supercooled liquid.

Our main goal in this section is to derive the relation, Eq. (1.1), between the temperature derivative of $x$ and that of the molar excess entropy $s_e$ of a supercooled liquid and the relation, Eq. (1.2), between its viscosity and $x$.

### A. The molar excess volume

The molar volume $v$ of the supercooled liquid at temperature $T$ is a well-defined macroscopic quantity so that we can operationally define the molar excess volume of the supercooled liquid by

$$v_e(T) \equiv v(T) - v_{crystal}(T), \tag{3.1}$$

where $v_{crystal}$ is the molar volume of the equilibrium crystal at the same temperature $T$. Close to $T_g$, we assume that the supercooled liquid is a mixture of the solid-like and liquid-like micro regions so that $v_e$ satisfies

$$v_e(T) = \{1 - x(T)\}v_g + x(T)v_l = v_g + x(T)\Delta v, \tag{3.2}$$

where

$$v_g \equiv \lim_{T \to T_g^+} v_e(T) \qquad (3.3)$$

and $v_l$ is the molar excess volume of the liquid-like micro regions close to $T_g$ and is assumed to be independent of temperature over a narrow temperature range above $T_g$. As we assume that the molar volume of the solid-like micro regions is lower than that of the liquid-like micro regions, we assume $\Delta v \equiv v_l - v_g > 0$.

## B. The molar excess internal energy

The molar internal energy $u$ of the supercooled liquid at temperature $T$ is just the total energy per mole of its constituent atoms or molecules and is therefore a well-defined macroscopic quantity so that we can operationally define the molar excess internal energy of the supercooled liquid by

$$u_e(T) \equiv u(T) - u_{crystal}(T), \qquad (3.4)$$

where $u_{crystal}$ is the molar internal energy of the equilibrium crystal at the same temperature $T$. Close to $T_g$, we assume that $u_e$ satisfies

$$u_e(T) = \{1 - x(T)\}u_g + x(T)u_l = u_g + x(T)\Delta u, \qquad (3.5)$$

where

$$u_g \equiv \lim_{T \to T_g^+} u_e(T) \qquad (3.6)$$

and $u_l$ is the molar excess internal energy of the liquid-like micro regions close $T_g$ and is assumed to be independent of temperature over a narrow temperature range above $T_g$. We assume that the molar internal energy of the solid-like micro regions is not higher than that of the liquid-like micro regions so that $\Delta u \equiv u_l - u_g \geq 0$.

## C. The molar excess enthalpy

We can operationally define the molar enthalpy of the supercooled liquid at temperature $T$ by

$$h(T) \equiv u(T) + Pv(T) \qquad (3.7)$$

so that we can define the molar excess enthalpy of the supercooled liquid by

$$h_e(T) \equiv h(T) - h_{\text{crystal}}(T), \qquad (3.8)$$

where $h_{\text{crystal}}(T) \equiv u_{\text{crystal}}(T) + Pv_{\text{crystal}}(T)$. $h_e$ then satisfies

$$h_e(T) = u_e(T) + Pv_e(T) = h_g + x(T)\Delta h, \qquad (3.9)$$

where $h_g \equiv u_g + Pv_g$ while $\Delta h$ defined by $\Delta h \equiv \Delta u + P\Delta v$ is the molar excess enthalpy difference between the solid-like micro regions and the liquid-like micro regions close to $T_g$ and is assumed to be almost independent of temperature over a narrow temperature range above $T_g$. We assume $\Delta v > 0$ and $\Delta u \geq 0$ so that $\Delta h > 0$.

Eq. (3.9) together with $\Delta h > 0$ is the basic assumption of our macroscopic model.

## D. The molar enthalpy and the molar heat capacity at constant pressure

As long as the supercooled liquid is going through a very slow infinitesimal process so that the pressure inside the supercooled liquid is uniform on the macroscopic level, we can use the law of conservation of energy together with the expression for a mechanical work $\delta W$ done on the liquid in terms of the pressure $P$ and a volume change $dV$ so that a change $dU$ in the internal energy of the supercooled liquid satisfies

$$dU = \delta Q + \delta W = \delta Q - PdV \, , \qquad\qquad (3.10)$$

where $\delta W = -PdV$ and $\delta Q$ is the heat transferred into the liquid during the process. $\delta Q$ then satisfies

$$\delta Q = dU + PdV \, , \qquad\qquad (3.11)$$

where $n$ is the mole number of the supercooled liquid.

We then operationally define the molar heat capacity at constant pressure $c_P$ of the supercooled liquid by

$$\delta Q_P = nc_P dT \, , \qquad\qquad (3.12)$$

where $\delta Q_P$ is the heat transferred into the supercooled liquid during an infinitesimal isobaric process at $P$. As $\delta Q_P$ also satisfies

$$\delta Q_P = d(U + PV) = nd(u + Pv) = ndh = n\left(\frac{\partial h}{\partial T}\right)_P dT \, , \qquad\qquad (3.13)$$

we find that $c_P$ is equal to the temperature derivative of the molar enthalpy:

$$c_P = \left(\frac{\partial h}{\partial T}\right)_P . \qquad\qquad (3.14)$$

**E. The molar excess enthalpy and the molar excess heat capacity at constant pressure**

We operationally define the molar excess heat capacity at constant pressure $c_P^{\mathrm{e}}$ for the supercooled liquid by

$$c_P^{\mathrm{e}} \equiv c_P - c_P^{(\mathrm{crystal})}, \qquad (3.15)$$

where

$$c_P^{(\mathrm{crystal})} = \left( \frac{\partial h_{\mathrm{crystal}}}{\partial T} \right)_P \qquad (3.16)$$

is the molar heat capacity at constant pressure of the equilibrium crystal. Using Eq. (3.14) and Eq. (3.16) in Eq. (3.15), we find $c_P^{\mathrm{e}}$ to satisfy

$$c_P^{\mathrm{e}} = \left( \frac{\partial h_{\mathrm{e}}}{\partial T} \right)_P = \Delta h \left( \frac{\partial x}{\partial T} \right)_P, \qquad (3.17)$$

where we have also used Eq. (3.9). $c_P^{\mathrm{e}}$ thus controls the temperature dependence of $x$.

## F. The molar excess entropy

We operationally define the molar excess entropy of the supercooled liquid at temperature $T$ by

$$s_{\mathrm{e}}(T) \equiv s(T) - s_{\mathrm{crystal}}(T), \qquad (3.18)$$

where $s$ is operationally defined by

$$s(T) \equiv s_{\mathrm{liquid}}(T_{\mathrm{m}}) - \int_T^{T_{\mathrm{m}}} dT' \frac{c_P(T')}{T'}, \qquad (3.19)$$

where $s_{\text{liquid}}(T_{\text{m}})$ is the molar entropy of the equilibrium liquid state at $T_{\text{m}}$, the melting temperature of the equilibrium crystal. The molar entropy $s_{\text{crystal}}$ of the equilibrium crystal also at $T$ can be estimated by

$$s_{\text{crystal}}(T) = s_{\text{liquid}}(T_{\text{m}}) - \frac{l_{\text{fus}}}{T_{\text{m}}} - \int_{T}^{T_{\text{m}}} dT' \frac{c_P^{(\text{crystal})}(T')}{T'}, \tag{3.20}$$

where $l_{\text{fus}}$ is the molar latent heat of fusion of the equilibrium crystal. The molar excess entropy then satisfies

$$s_{\text{e}}(T) = s_{\text{e}}(T_{\text{m}}) - \int_{T}^{T_{\text{m}}} dT' \frac{c_P^{\text{e}}(T')}{T'} = s_{\text{e}}(T_{\text{g}}) + \int_{T_{\text{g}}}^{T} dT' \frac{c_P^{\text{e}}(T')}{T'}, \tag{3.21}$$

where we have used

$$\frac{l_{\text{fus}}}{T_{\text{m}}} = s_{\text{liquid}}(T_{\text{m}}) - s_{\text{crystal}}(T_{\text{m}}) = s_{\text{e}}(T_{\text{m}}). \tag{3.22}$$

## G. The mole fraction $x$ of the liquid-like micro regions and the molar excess entropy

Eq. (3.21) implies

$$c_P^{\text{ex}} = T \left( \frac{\partial s_{\text{ex}}}{\partial T} \right)_P. \tag{3.23}$$

Using this equation and Eq. (3.17), we find the temperature dependence of the mole fraction $x$ of the liquid-like micro regions to be directly related to that of the molar excess entropy $s_{\text{e}}$ by

$$\left( \frac{\partial x}{\partial T} \right)_P = \frac{T}{\Delta h} \left( \frac{\partial s_{\text{e}}}{\partial T} \right)_P. \tag{3.24}$$

## H. The mole fraction $x$ of the liquid-like micro regions and the viscosity

Close to $T_g$, we assume that an activated motion of the solid-like micro regions controls the viscosity $\eta$ of the supercooled liquid. We also assume that this activated motion is a collective motion involving practically all of the solid-like micro-regions so that the molar activation free energy $\Delta g_a$ for the activated motion is proportional to the mole fraction, $1-x$, of the solid-like micro regions. We then obtain

$$\eta = \eta_0 \exp\left(\frac{\Delta g_a}{RT}\right) = \eta_0 \exp\left\lfloor \frac{\Delta g_0(1-x)}{RT}\right\rfloor, \qquad (3.25)$$

where $R$ is the universal gas constant and $\Delta g_0$ is a constant specific to a particular liquid.

Eq. (3.24) and Eq. (3.25) are the key results of our model and they show that the temperature dependence of the viscosity $\eta$ is connected to that of the molar excess entropy $s_e$ through the temperature dependence of the mole fraction $x$.

Note that our assumption, $\Delta g_a = \Delta g_0(1-x)$, is more restrictive than a general linearization of $\Delta g_a$, $\Delta g_a = \Delta g_0 - \Delta g_1 x$, with an extra parameter $\Delta g_1$.

Using Eq. (3.24) and Eq. (3.25), we also find the molar excess heat capacity at constant pressure $c_P^e$ of the liquid to be related to the viscosity $\eta$ by

$$c_P^e = T\left(\frac{\partial s_e}{\partial T}\right)_P = \Delta h\left(\frac{\partial x}{\partial T}\right)_P = -\frac{R\Delta h}{\Delta g_0}\left\lfloor\frac{\partial}{\partial T}\left\{T\ln\left(\frac{\eta}{\eta_0}\right)\right\}\right\rfloor_P. \qquad (3.26)$$

We can therefore test our model for a particular glass former by finding whether its experimental data for $c_P^e$ and $\eta$ are consistent with this equation with two adjustable parameters, $\eta_0$ and $R\Delta h/\Delta g_0$.

## IV. APPLICATION OF THE MODEL TO A CLASS OF GLASS FORMERS

In this section, we apply our model to a class of glass formers for which the molar excess entropy $s_e$ at temperatures close to $T_g$ is well approximated by the following equation[3]:

$$\frac{s_e(T)}{s_\infty} = 1 - \frac{T_K}{T} \qquad \text{(for } T \geq T_g > T_K\text{),} \qquad (4.1)$$

where both $s_\infty$ and $T_K$, the Kauzmann temperature, are constants specific to a particular liquid.

For this class of glass formers, we will also assume that $T_g^*$ is very close to $T_g$ so that we can approximate $T_g^*$ by $T_g$. This is not an entirely unreasonable assumption since for this class, we find $T_K \sim 0.8 T_g$ and $T_g^*$ must be between $T_g$ and $T_K$.

### A. The mole fraction $x$ of the liquid-like micro regions and the molar excess entropy

Using Eq. (4.1) in Eq. (3.24), we obtain

$$\left(\frac{\partial x}{\partial T}\right)_P = \frac{T}{\Delta h}\left(\frac{\partial s_e}{\partial T}\right)_P = \frac{s_\infty T_K}{\Delta h T}. \qquad (4.2)$$

Integrating the both sides of this equation and using $x(T_g) \equiv \lim_{T \to T_g^*} x(T) = 0$, we find

$$x = a \ln\left(\frac{T}{T_g}\right), \qquad (4.3)$$

where the parameter $a$ is defined by

$$a \equiv \frac{s_\infty T_K}{\Delta h}. \qquad (4.4)$$

Note that Eq. (4.3) holds only close to $T_g$ so that, for example, $a \ln\left(T_m/T_g\right) \neq 1$.

**B. The mole fraction $x$ of the liquid-like micro regions and the viscosity**

Using Eq. (4.3) in Eq. (3.25), we then obtain the viscosity $\eta$ of the supercooled liquid as

$$\eta = \eta_0 \exp\left[\frac{\Delta g_0(1-x)}{RT}\right] = \eta_0 \exp\left[\frac{\Delta g_0}{RT}\left\{1 - a\ln\left(\frac{T}{T_g}\right)\right\}\right], \qquad (4.5)$$

Note that this expression for $\eta$ is valid only for temperatures very close to $T_g$.

For temperatures very close to $T_g$, we can approximate the mole fraction $x$ by

$$x = a\ln\left(\frac{T}{T_g}\right) = -a\ln\left(\frac{T_g}{T}\right) = -a\ln\left[1 - \left(1 - \frac{T_g}{T}\right)\right] \cong a\left(1 - \frac{T_g}{T}\right), \qquad (4.6)$$

where we have used $\ln(1-y) \cong -y$ for $0 < y << 1$. We can then approximate $\Delta g_0(1-x)$ by

$$\Delta g_0(1-x) \cong \Delta g_0\left\{1 - a\left(1 - \frac{T_g}{T}\right)\right\} \cong \frac{\Delta g_0}{1 + a\left(1 - \frac{T_g}{T}\right)} = \frac{\Delta g_0 T}{(a+1)\left(T - \frac{a}{a+1}T_g\right)}, \qquad (4.7)$$

where we have used $1 - y \cong 1/(1+y)$ for $0 < y << 1$. For temperatures very close to $T_g$, the viscosity is therefore well approximated by the following Vogel-Fulcher-Tamman equation:

$$\eta = \eta_0 \exp\left(\frac{D T_{\text{VFT}}}{T - T_{\text{VFT}}}\right) = \eta_0 \exp\left(\frac{D}{T/T_{\text{VFT}} - 1}\right), \qquad (4.8)$$

where

$$T_{\mathrm{VFT}} = \frac{a}{a+1} T_{\mathrm{g}} < T_{\mathrm{g}} \qquad (4.9)$$

and

$$D = \frac{\Delta g_0}{R T_{\mathrm{VFT}}(a+1)} = \frac{\Delta g_0}{R T_{\mathrm{g}} a}. \qquad (4.10)$$

## C. Three parameters in our model

Using Eq. (4.9), we can estimate the parameter $a$ by

$$a = \frac{T_{\mathrm{VFT}}}{T_{\mathrm{g}} - T_{\mathrm{VFT}}} = \frac{1}{T_{\mathrm{g}}/T_{\mathrm{VFT}} - 1}. \qquad (4.11)$$

The parameter $a$ thus depends only on $T_{\mathrm{g}}/T_{\mathrm{VFT}}$. The larger the ratio $T_{\mathrm{g}}/T_{\mathrm{VFT}}$ is, the smaller the parameter $a$ is. As the parameter $a$ approaches zero, $T_{\mathrm{VFT}}$ also approaches zero so that the Vogel-Fulcher Tamman equation approaches the Arrhenius equation, which implies that the parameter $a$ is a measure for the fragility of a supercooled liquid: the smaller the parameter $a$ is, the less fragile the supercooled liquid is.

According to Eq. (4.3), the larger the parameter $a$ is, the more rapidly the mole fraction $x$ of the liquid-like micro regions increases as the temperature of a supercooled liquid is increased. This implies that the solid-like micro regions inside a fragile glass former are more easily destroyed by an increase in the temperature than those in a less fragile glass former. In other words, the solid-like micro regions inside a fragile glass former are "fragile" against an increase in the temperature. The parameter $a$ is therefore a measure for the fragility of the solid-like micro regions inside a supercooled liquid very close to $T_{\mathrm{g}}$.

Using Eq. (4.11), we can also express the mole fraction $x$ in terms of $T$, $T_{\mathrm{g}}$, and $T_{\mathrm{g}}/T_{\mathrm{VFT}}$:

$$x = a \ln\left(\frac{T}{T_g}\right) = \left(\frac{1}{T_g/T_{VFT} - 1}\right) \ln\left(\frac{T}{T_g}\right). \tag{4.12}$$

Using Eq. (4.4) and Eq. (4.11), we can also estimate $\Delta h$, the second parameter in our model, by

$$\Delta h = \frac{s_\infty T_K}{a} = s_\infty T_K \left(\frac{T_g}{T_{VFT}} - 1\right). \tag{4.13}$$

Using Eq. (4.10), we can estimate $\Delta g_0 \big/ \left(R T_g\right)$, the third parameter in our model, by

$$\frac{\Delta g_0}{R T_g} = Da = \frac{D T_{VFT}}{T_g - T_{VFT}} = \frac{D}{T_g/T_{VFT} - 1} = \ln\left\lfloor \frac{\eta(T_g)}{\eta_0} \right\rfloor \tag{4.14}$$

so that $\Delta g_0 \big/ \left(R T_g\right)$ depends on $D$ and $T_g/T_{VFT}$ or $\eta(T_g)$ and $\eta_0$. If we choose $\eta(T_g) = 10^{13}$ P, then $\Delta g_0 \big/ \left(R T_g\right)$ depends only on $\eta_0$.

## D. The steepness index $m$ and the strength parameter $D$

As a measure for the fragility of a supercooled liquid, we often use the steepest index $m$, which depends on the parameter $a$ and $\Delta g_0 \big/ \left(R T_g\right)$ as

$$m \equiv \frac{d \log_{10}(\eta/\text{s})}{d\left(T_g/T\right)}\bigg|_{T=T_g} = \frac{\log_{10}(e) D T_{VFT} T_g}{\left(T_g - T_{VFT}\right)^2} = \log_{10}(e) D \left(\frac{1}{T_g/T_{VFT} - 1}\right)^2 \frac{T_g}{T_{VFT}}$$

$$= \log_{10}(e)\left(\frac{\Delta g_0}{R T_g a}\right) a^2 \frac{a+1}{a} = \frac{\log_{10}(e) \Delta g_0 (a+1)}{R T_g}. \tag{4.15}$$

Note also that we can estimate $m$ directly from $D$ and $T_g/T_{VFT}$.

Another frequently used measure for the fragility is the strength parameter $D$ that also depends on the parameter $a$ and $\Delta g_0 / (RT_g)$ as shown in Eq. (4.10). For three glass formers listed in Table I, we find $\ln\left[\eta\left(T_g\right) / \eta_0\right] = \Delta g_0 / \left(RT_g\right) \sim 40$, which follows from $\eta\left(T_g\right) \sim 10^{13}$ P and $\eta_0 \sim 10^{-4}$ P, a common high temperature limiting value for the viscosity, so that $m$ is roughly proportional to $a+1$ and $D$ is roughly inversely proportional to $a$.

As mentioned in Sec. IV.C, the parameter $a$ is a measure for the fragility of the solid-like micro regions inside a supercooled liquid very close to $T_g$ and it appears to possess a more transparent physical meaning compared with $m$ and $D$.

As $m/D$ depends only on $T_g / T_{\text{VFT}}$, it depends only on $a$:

$$\frac{m}{D} = \frac{\log_{10}(e)\left(T_g / T_{\text{VFT}}\right)}{\left(T_g / T_{\text{VFT}} - 1\right)^2} = \log_{10}(e)a^2\left(\frac{a+1}{a}\right) = \log_{10}(e)a(a+1). \qquad (4.16)$$

This suggests that $m/D$ can be used also as a measure for the fragility of a supercooled liquid.

**E. $T_{\text{K}}$ and $T_{\text{VFT}}$**

In our model, $T_{\text{K}}$ and $T_{\text{VFT}}$ do not need to be equal or close to each other. Using Eq. (4.4), we can express $s_\infty$ as

$$s_\infty = \frac{a\Delta h}{T_{\text{K}}}. \qquad (4.17)$$

Using this in Eq. (4.1), we obtain

$$s_{\text{ex}}\left(T_g\right) = \frac{a\Delta h}{T_{\text{K}}}\left(1 - \frac{T_{\text{K}}}{T_g}\right), \qquad (4.18)$$

so that

$$T_{\text{K}} = \frac{a}{a + T_{\text{g}} s_{\text{ex}}\left(T_{\text{g}}\right)\!\big/\Delta h} T_{\text{g}}. \qquad (4.19)$$

Comparing this equation with Eq. (4.9), we find that if $T_{\text{g}} s_{\text{ex}}\left(T_{\text{g}}\right)\!\big/\Delta h \approx 1$, then $T_{\text{K}} \approx T_{\text{VFT}}$ whereas if $T_{\text{g}} s_{\text{ex}}\left(T_{\text{g}}\right)\!\big/\Delta h < 1$, then $T_{\text{K}} > T_{\text{VFT}}$ (see Table I for examples).

## V. THE THREE PARAMETERS IN OUR MODEL FROM EXPERIMENTAL DATA

Once we estimate the values for $T_{\text{g}}$, $s_{\infty}$, $T_{\text{K}}$, $T_{\text{VFT}}$, and $D$ from the experimental data for a particular glass former in the class discussed in Sec. IV, we can estimate the values for the three parameters, $a$, $\Delta h$, and $\Delta g_0\big/\left(RT_{\text{g}}\right)$, as listed in Table I. Keep in mind that we are assuming that the reported value of $T_{\text{g}}$ for a particular glass former can be used as a good approximation for the lowest glass transition temperature for the slowest possible cooling rate.

To estimate the values of the three parameters for a particular glass former, we need its values of three quantities, $T_{\text{g}}\big/T_{\text{VFT}}$, $s_{\infty}T_{\text{K}}$, and $D$. Using Eq. (4.11), we can estimate the parameter $a$ from $T_{\text{g}}\big/T_{\text{VFT}}$. Using Eq. (4.13), we can estimate $\Delta h$ from $s_{\infty}T_{\text{K}}$ and $T_{\text{g}}\big/T_{\text{VFT}}$. Using Eq. (4.14), we can estimate $\Delta g_0\big/\left(RT_{\text{g}}\right)$ from $D$ and $T_{\text{g}}\big/T_{\text{VFT}}$.

We can also estimate $s_{\text{e}}\left(T_{\text{g}}\right)$ from $T_{\text{g}}$, $T_{\text{K}}$, and $s_{\infty}$ using Eq. (4.1), $m$ from $D$ and $T_{\text{g}}\big/T_{\text{VFT}}$ (Eq. (4.15)), and $m\big/D$ from $T_{\text{g}}\big/T_{\text{VFT}}$ (Eq. (4.16)).

In Table I, we list the estimated values for these macroscopic quantities for three glass formers, $o$-terphenyl, 3-bromopentane, and $Pd_{40}Ni_{40}P_{20}$. Among these three, $o$-terphenyl is the most fragile while $Pd_{40}Ni_{40}P_{20}$ is the least fragile.

For $o$-terphenyl and 3-bromopentane in Table I, we used the values for $T_{\text{g}}$, $s_{\infty}$, and $T_{\text{K}}$ estimated by Richert and Angell[3], who used the calorimetric data for $o$-terphenyl obtained by Chang and Bestul[7] as well as the data for $c_P$ and $c_P^{(\text{crystal})}$ for 3-bromopentane obtained by S. Takahara, O. Yamamuro, and T. Matsuo[8].

For $Pd_{40}Ni_{40}P_{20}$, we used the value for $T_g$ obtained by Wilde *et al.*[9]. By comparing their expression for $c_P^e$ with the following linear approximation just above $T_g$,

$$c_P^e = T\left(\frac{\partial s_e}{\partial T}\right)_P = \frac{s_\infty T_K}{T} \cong \frac{s_\infty T_K}{T_g}\left(2 - \frac{T}{T_g}\right), \qquad (5.1)$$

we obtained $s_\infty T_K / T_g = 21 \text{ J}/\left(K^2 \cdot mol\right)$. Using their expression for $c_P^e$ for the temperature range $T_g \le T \le T_m = 884$ K and their value for $l_{fus}$, we obtained $s_e\left(T_g\right) = 3.6 \text{ J}/(K \cdot mol)$. Using Eq. (4.1), we then found $s_\infty = 25 \text{ J}/(K \cdot mol)$ and using Eq. (4.19), we also obtained $T_K = 490$ K, which is close to their estimate of $T_K = 500 \pm 5$ K.

The estimated values for $T_{VFT}$ and $D$ are also listed in Table I. For *o*-terphenyl and 3-bromopentane, we used the values for $T_{VFT}$ and $D$ estimated by Richert and Angell[3], who used their data for the relaxation time for *o*-terphenyl as well as the data for the relaxation time for 3-bromopentane obtained by Berberian and Cole[10]. For $Pd_{40}Ni_{40}P_{20}$, we used the values for $T_{VFT}$ and $D$ obtained by Kawamura and Inoue[11] from their data for the viscosity.

It is interesting to note that the values of the molar excess entropy at $T_g$ for *o*-terphenyl and 3-bromopentane, for which we find $T_K \cong T_{VFT}$, are rather close at $s_e\left(T_g\right) \cong 23 \text{ J}/(K \cdot mol)$ while they are much higher than $s_e\left(T_g\right) \cong 3.6 \text{ J}/(K \cdot mol)$ for $Pd_{40}Ni_{40}P_{20}$, for which we find $T_K > T_{VFT}$. For another fragile glass former, 2-methyltetrahydrofuran[3], for which we also find $T_K \cong T_{VFT}$, the value of $s_e\left(T_g\right)$ is also roughly $23 \text{ J}/(K \cdot mol)$.

We also find that the values of $\Delta h/\left(RT_g\right)$ for *o*-terphenyl, 3-bromopentane, and 2-methyltetrahydrofuran are also rather close at $\Delta h/\left(RT_g\right) \cong 3$ while they are much higher than $\Delta h/\left(RT_g\right) \cong 1$ for $Pd_{40}Ni_{40}P_{20}$. This is not surprising since using Eq. (4.13), we find

$$\frac{\Delta h}{RT_g} = \frac{s_\infty T_K}{RT_g}\left(\frac{T_g}{T_{VFT}} - 1\right) = \frac{s_\infty}{R}\left(\frac{T_K}{T_{VFT}} - \frac{T_K}{T_g}\right) = \left\{\frac{s_e(T_g)}{R}\right\}\frac{\dfrac{T_K}{T_{VFT}} - \dfrac{T_K}{T_g}}{1 - \dfrac{T_K}{T_g}} \qquad (5.2)$$

so that if the values of $s_e(T_g)$ are close for two fragile glass formers that satisfy $T_K \cong T_{VFT}$, then the values of $\Delta h/(RT_g)$ for them should be also close.

As pointed out in Sec. IV.E, for *o*-terphenyl and 3-bromopentane, we find $T_g s_e(T_g)/\Delta h \cong 1$, which is consistent with $T_K \cong T_{VFT}$ found for these glass formers whereas for $Pd_{40}Ni_{40}P_{20}$, we find $T_g s_e(T_g)/\Delta h = 0.37$, which is consistent with $T_K > T_{VFT}$ found for this glass former.

## VI. TESTING OUR ASSUMPTION ABOUT THE DENSITIES OF THE SOLID-LIKE AND LIQUID-LIKE MICRO REGIONS BY MOLECULAR DYNAMICS SIMULATIONS

Our model is motivated by the physical picture of deeply supercooled liquids as mixtures of solid-like and liquid-like micro regions. We assume that these solid-like micro regions have, on average, a slightly higher density than the liquid-like micro regions so that, on the macroscopic level, the molar volume of the solid-like micro regions is lower than that of the liquid-like micro regions. To test this assumption, we may run constant-temperature constant-pressure molecular dynamics simulations of a model liquid at temperatures close to $T_g$ and examine the distribution $p(v_i)$ of Voronoi cell volumes for the liquid.

For example, Yonezawa[12] calculated $p(v_i)$ for a single component Lennard-Jones system above and below $T_g$. Above $T_g$, $p(v_i)$ is a sum of two Gaussian distributions (see the frame for $t = 2000$ in Fig. 41 in Ref. 12). The one centered at a lower volume must correspond to the atoms in the solid-like micro regions while the other centered at a higher volume must correspond to the atoms in the liquid-like micro regions. Below $T_g$, $p(v_i)$ cannot be fit by a sum of two Gaussian distributions (see the frame for $t = 25000$ in Fig. 42 in Ref. 12) and must represent a randomly connected network of solid-like micro regions.

More detailed examinations of $p(v_i)$ for other model liquids should help us further test our assumption. For example, from the relative statistical weight of the two Gaussian distributions, we can estimate the mole fraction $x$ of the liquid-like micro regions to check whether it can be fit by Eq. (4.3). This fit then yields a value for the parameter $a$, which we can compare with the value of $a$ estimated by Eq. (4.11).

## VII. CONCLUSIONS

In this article, for a deeply supercooled liquid near its glass transition temperature, we have suggested a possible way to connect the temperature dependence of its molar excess entropy and that of its viscosity by constructing a macroscopic model, where the deeply supercooled liquid is assumed to be a mixture of solid-like and liquid-like micro regions. In our model, the temperature dependence of the viscosity is connected to that of the molar excess entropy $s_e$ through the temperature dependence of the mole fraction $x$.

In our model, we assume that the mole fraction $x$ of the liquid-like micro regions tends to zero as the temperature $T$ of the liquid is decreased and extrapolated to a temperature $T_g^*$, which we assume to be below but close to the lowest glass transition temperature $T_g$ attainable with the slowest possible cooling rate for the liquid. Without referring to any specific microscopic nature of the solid-like and liquid-like micro regions, we also assume that near $T_g$, the molar enthalpy of the solid-like micro regions is lower than that of the liquid-like micro regions.

We have shown that the temperature dependence of $x$ is directly related to that of the molar excess entropy. Close to $T_g$, we assume that an activated motion of the solid-like micro regions controls the viscosity and that this activated motion is a collective motion involving practically all of the solid-like micro-regions so that the molar activation free energy $\Delta g_a$ for the activated motion is proportional to the mole fraction, $1 - x$, of the solid-like micro regions. The temperature dependence of the viscosity is thus connected to that of the molar excess entropy $s_e$ through the temperature dependence of the mole fraction $x$.

As an example, we have applied our model to a class of glass formers for which $s_e$ at temperatures near $T_g$ is well approximated by $s_e \propto 1 - T_K/T$ with $T_K < T_g \equiv T_g^*$ and find their viscosities to be well approximated by the Vogel-Fulcher-Tamman equation for temperatures very close to $T_g$.

We have also found that a parameter $a$ appearing in the temperature dependence of $x$ for a glass former in this class is a measure for its fragility. As this class includes both fragile and strong glass formers, our model applies to both fragile and strong glass formers.

Our macroscopic model applied to this class of glass formers contains three parameters: the parameter $a$, the molar excess enthalpy difference $\Delta h$ between the solid-like micro regions and the liquid-like micro regions close to $T_g$, and $\Delta g_0 / (R T_g)$, where $R$ is the universal gas constant and $\Delta g_0$ is the molar activation free energy for the activated motion of the solid-like micro regions at $T_g$. We have estimated the values of these parameters for three glass formers in this class: $o$-terphenyl, 3-bromopentane, and $Pd_{40}Ni_{40}P_{20}$, which is the least fragile among the three.

Finally, we have suggested a way to test our assumption about the solid-like and liquid-like micro regions using molecular dynamics simulations of model liquids.

**ACKNOWLEDGMENTS**

I wish to thank Michele Bock for constant support and encouragement and Richard F. Martin, Jr. and other members of the physics department at Illinois State University for creating a supportive academic environment.

TABLE I  Experimental values for $T_g$, $T_K$, $s_\infty$, $T_{VFT}$ and $D$ for three glass formers (for references for these values, see the text) as well as macroscopic quantities in our model estimated from these values. $a$ from $T_g/T_{VFT}$ (Eq. (4.11)). $\Delta h$ from $s_\infty T_K$ and $T_g/T_{VFT}$ (Eq. (4.13)), $s_c(T_g)$ from $T_g$, $T_K$, and $s_\infty$ (Eq. (4.1)), $\Delta g_0/(RT_g)$ from $D$ and $T_g/T_{VFT}$ (Eq. (4.14)). $m$ from $D$ and $T_g/T_{VFT}$ (Eq. (4.15)). $m/D$ from $T_g/T_{VFT}$ (Eq. (4.16)).

| Substance | $o$-terphenyl | 3-bromopentane | $Pd_{40}Ni_{40}P_{20}$ |
|---|---|---|---|
| $T_g$ (K) | 246 | 107 | 570 |
| $T_K$ (K) | 204.2 | 82.5 | 490 |
| $s_\infty$ (J/(K · mol)) | 137.4 | 99.0 | 25 |
| $T_{VFT}$ (K) | 202.4 | 82.9 | 390 |
| $D$ | 7.78 | 10.4 | 18.1 |
| $a$ | 4.65 | 3.44 | 2.2 |
| $\Delta h$ (J/mol) | 6030 | 2370 | 5600 |
| $\Delta h/(RT_g)$ | 2.95 | 2.66 | 1.2 |
| $s_c(T_g)$ (J/(K · mol)) | 23.4 | 22.7 | 3.6 |
| $T_g s_c(T_g)/\Delta h$ | 0.955 | 1.02 | 0.37 |
| $\Delta g_0/(RT_g)$ | 36.2 | 35.8 | 40 |
| $m$ | 88.8 | 69.0 | 56 |
| $m/D$ | 11.4 | 6.63 | 3.1 |